\documentclass[aps,prl,twocolumn,groupedaddress]{revtex4}

\begin{document}

\title{"Continuum versus Discrete" Physics: brief note on \\ Endo- and exo-observer of a physical system.}

\author{David Vernette}
\author{Michele Caponigro}

\affiliation{}

\emph{\date{\today}}
\\
\\
\begin{abstract}
We argue about the following concepts:(i)introduction of endo
and exo-observer of a physical system (ii) possible relation between endo, exo-observer and
continuum/discrete nature of the same system (iii)the distinction about two categories of observers do not help
to solve the basic nature (continuum/discrete)of the system considered.
We argue that \emph{discrete} math tool utilized to
describe behavior of a physical system is to ascribe to the exo-observer description.
In other words, discreteness nature of a system is due at the
presence of an external observer description, the observation act.
Vice versa, we retain, that endo-observer
description give us a continuum nature of the system but only in a Platonic realm.
Finally, we retain that a real theory of the state of matter could to build
a real independence from the observer.
These brief considerations open a discussion on the
foundations of discrete math tools and their relation with the observation act,
included the possible extension to foundations of the quantum mechanics.
\end{abstract}

\maketitle

\section{Introduction: role of the observer}

Recently many works have been dedicated to informational approaches to the
physics, especially in the field of quantum mechanics (Fuchs, Peres,
Zeilinger\cite{Fu}). We are interested to analyzing this approach respect the problem of continuum/dicrete nature
of physical system.
First, we retain that:
\begin{itemize}
\item 1)The informational approach is not a prerogative of quantum mechanics.The structure of informational approach
is applicable in many fields of physics.
\item 2)The approach is based on: i) a subjective interpretation of probability
ii)the equivalence between physical reality and our knowledge of
physical reality,this position is supported through the irreducibly random nature of individual quantum events.
\end{itemize}

As we have seen, the approach is centered on the fundamental role of the observer's description.
We retain that the presence or apparent absence of the observer is
an important step to establish the nature of physical system.
What we intend with observer? Which is the nature of the "relation" with the correspondent physical system?\\

We adopt the following general assumptions:
\begin{itemize}
\item 1)We call \textbf{exo-observer system}: a physical system which is mathematically dependent from an external observer
description(e.g. math description of the system depend by external observer).
\item 2)We call \textbf{endo-observer system}: a physical system which is not mathematically dependent by any external
observer description.
\end{itemize}

We think that above assumptions are not sufficient to distinguish the basic
nature (continuun/discrete) of a physical system. We will analyze this point next section.
Of course, the ideal hypothesis is that exist a metalanguage where any endo/exo-observer
can be formulated. We do not see (until now) the existence of a (neutral)
metalanguage. We will argue that the possible existence of a
metalanguage must be linked with a real theory of the state of matter.
For this reason, our brief analysis will be more general and will diverge from
Primas\cite{Pri} position: \emph{Inescapably, the study of the
endo/exo-dichotomy must be conducted in a metalanguage. The
language in which any endo/exo-theory can ever be formulated is
neither part of the [...]} endo and exo-observer description.
We agree,instead,with the following Primas\cite{Pri}definition, which he
calls "the observer-free"(quantum endophysics):\emph{the study of the Platonic heaven,
the realm of non-spatial, non-mental, timeless but nevertheless real
entities. \textbf{According Primas}: A hard-boiled positivist may have
difficulties to appreciate such a quantum endophysics since it
refers by definition to some kind of a Platonic universe, and not
to empirical facts. The aim of the study of quantum endophysics
\textbf{is not the hope to find the "true and real" values for all
endophysical observables}(these do not exist!) but to formulate
universally valid natural laws endophysically in view of
theoretical derivations of operationally meaningful exophysical
descriptions.}

\section{Continuum vs Discrete: Platonic Realm vs Information}

First, we do not retain a negative element the "subjectivity" in a description of a physical system.
We argue simple that the "exo-observer act" introduce a discrete behavior of the physical system.
Vice versa, we retain, that endo-observer description give us a "continuum" nature of the system but only
in a \textbf{Platonic realm}. The tension between discrete and continuous aspects
seem without a definitive solution.
As we know, from ancient philosophy the question of the nature, for instance, of space
(atomicity vs.infinite divisibility) and its constituting
elements has played a great part in metaphysical studies. This
question raised some problems that still make up interesting
research topics not only in philosophy but in mathematics and
cognitive sciences as well. What does constitute space? Here, our
intentions are not to get into philosophical analysis, but rather
to try to evidence the fundamental role of the observer in the
description of the system.

Discrete (respectively continuous) can be related to: time, real space, phase
space etc. This analysis could be done under different
levels: a) computational techniques (i.e.informational approach) b)
nature of the phenomenon (i.e. quantum nature).
The meaning of "continuous", for instance, seem less ambiguous in physics than in mathematics.

Which are the implications of two observer's category?
\begin{itemize}
\item A)\textbf{endo-observer} $\Rightarrow$This hidden observer usually is introduced under form of "physical laws"
(Platonic Observation). For instance, the Shr\"{o}dinger's equation:
$i\hbar\frac{\partial\Psi}{\partial t}$= H$\Psi$)$\rightarrow$ is an example of "\textbf{Platonic observation}".
\item B)\textbf{exo-observer}$\Rightarrow$This case is introduced under form of "information" (Subjective
Observation). For instance, the collapse of wavefunction
$\rightarrow$is an example of "\textbf{Subjective observation}".
\end{itemize}

We argue that both categories of observers do not solve the basic nature of continuum/discrete dichotomy of the physical system.
How we can infer and to support these statements? Where are the theoretical reasons?
\begin{itemize}
\item 1) Until now, Exo-observer description implies$\rightarrow$\\a discrete nature of the physical systems
(included the unsuspicious classical systems). The prerogative of this description is the subjective interpretation of mathematical data.
\item  2) Until now, Endo-observer description implies$\rightarrow$ a "continuum" nature of the system \textbf{but} only in the Platonic
realm (The Math models and Physical Laws). In other words, we have not any real "continuum" definition of a physical system out of Platonic realm.
\end{itemize}

The following table summarize our brief considerations:\\
\\
\\
{\small
\begin{tabular}{l|l}
\textbf{\textsf{Endo-Observer} }& \textbf{\textsf{Exo-Observer}} \\
1.\emph{Physical Law} & 1.\emph{Information} \\
2."Continuum"(in Platonic realm)& 2.Discreteness(Idealism)\\
3.Universality & 3.Contextuality \\
4.Possible Ontic intepretation & 4.Epistemic interpretation\\
\end{tabular}}\\
\\
\\
\\

\section{Conclusion}
In conclusion, both approaches do not give us a definitive response on the continuum/discrete
nature of physical system. How we can go out?
We retain that a real \textbf{theory of the state of matter}
could to build a real independence from the observer, the metalanguage
auspicate from Primas.
\\
\\
\section{{\tiny  }}
{\footnotesize------------------\\ $\diamond$\emph{David Vernette}:Quantum Philosophy Theories
www.qpt.org.uk\\david.vernette@qpt.org.uk
\\
$\diamond$ \emph{Michele Caponigro}: University of Camerino, Physics
Department michele.caponigro@unicam.it}
\\
\\

\end{document}